# Martingales, Detrending Data, and the Efficient Market Hypothesis

Joseph L. McCauley[+], Kevin E. Bassler[++], and Gemunu H. Gunaratne[+++]

Physics Department
University of Houston
Houston, Tx. 77204-5005
jmccauley@uh.edu

[+]Senior Fellow
COBERA
Department of Economics
J.E.Cairnes Graduate School of Business and Public Policy
NUI Galway, Ireland

[++]Texas Center for Superconductivity
University of Houston
Houston, Texas 77204-5005

[+++]Institute of Fundamental Studies
Kandy, Sri Lanka

Key Words: Martingales, Markov processes, detrending, memory, stationary and nonstationary increments, correlations, efficient market hypothesis.

## Abstract

We discuss martingales, detrending data, and the efficient market hypothesis for stochastic processes $x(t)$ with arbitrary diffusion coefficients $D(x,t)$. Beginning with x-independent drift coefficients $R(t)$ we show that Martingale stochastic processes generate uncorrelated, generally *nonstationary* increments. Generally, a test for a martingale is therefore a test for uncorrelated increments. A detrended process with an x-dependent drift coefficient is generally not a martingale, and so we extend our analysis to include the class of (x,t)-dependent drift coefficients of interest in finance. We explain why martingales look Markovian at the level of both simple averages *and* 2-point correlations. And while a Markovian market has no memory to exploit and presumably cannot be beaten systematically, it has never been shown that martingale memory cannot be exploited in 3-point or higher correlations to beat the market. We generalize our Markov scaling solutions presented earlier, and also generalize the martingale formulation of the efficient market hypothesis (EMH) to include (x,t)-dependent drift in log returns. We also use the analysis of this paper to correct a misstatement of the 'fair game' condition in terms of serial correlations in Fama's paper on the EMH. We end with a discussion of Levy's characterization of Brownian motion and prove that an arbitrary martingale is topologically inequivalent to a Wiener process.

## 1. Introduction

Recently [1] we focused on the condition for long time correlations like and including fractional Brownian motion (fBm), which is stationarity of the increments in a stochastic process x(t) with variance nonlinear in the time. There, we derived the 2-point and 1-point densities including the transition density for fBm. We will point out below that there are nonMarkov systems where the pair correlations canot be distinguished from those of a Markov process, but time series with stationary increments (like fBm) exhibit long time memory that can be seen at the level of pair correlations: fBm cannot be mistaken for a Markov process at the 2-point level. We correspondingly emphasized that neither 1-point averages nor Hurst exponents can be used to identify the presence or absence of history-dependence in a time series, or to identify the underlying stochastic process (see [2] for the conclusion that an equation of motion for a 1-point density cannot be used to decide if a process is Markovian or not). In the same paper, we pointed out that the opposite class, systems with no memory at all (Markov processes) and with x-independent drift coefficients generate uncorrelated, typically nonstationary increments. The conclusions in [1] about Markov processes are more general than we realized at the time. Here, we generalize that work by focusing on martingales.

In applications to finance, by "x" we always mean $x(t)=\ln(p(t)/p_c)$ where $p(t)$ is a price at time t and $p_c$ is a reference price, the consensus price or 'value' [3]. The consensus price $p_c$ is simply the price that determines the peak of the 1-point returns density $f_1(x,t)$. The reason why log *increments* $x(t;T)=\ln p(t+T)/p(t)$ and price differences $\Delta p=p(t+T)-p(t)$ generally *cannot* be taken as 'good' variables describing a stochastic process (neither theoretically nor in data analysis) is explained below in part 4. It is impossible for a martingale, excepting the special case of a variance linear in the time t, to develop either stochastic dynamics or probability theory based on *increments* $x(t;T)$ or $\Delta p$, because if the increments are nonstationary, as they generally are, then the starting time t matters and consequently histograms derived empirically from time series assuming that the starting time doesn't matter exhibit 'significant artifacts' like fat tails and spurious Hurst exponents [3,4]. In contrast, in a system with long time autocorrelations (like fBm), the stationary increment $x(t;T)=x(t+T)-x(t)=x(T)$, 'in distribution', is a perfectly good variable. But real markets [4] are hard to beat and rule out increment autocorrelations.

Stated briefly, if increments are nonstationary then the 1-point density that describes the increments is not independent of t (see part 5 below). Next, we define the required underlying ideas.

## 2. Conditional expectations with memory

Imagine a collection of time series generated by an unknown stochastic process that we would like to discover via data analysis. Simple averages require only a 1-point density $f_1(x,t)$, e.g., $<x^n(t)>=\int x^n f_1(x,t)dx$. No dynamical process can be identified by specifying merely either the 1-point density or a scaling exponent [1]. Both conditioned and unconditioned two-point correlations, e.g. $<x(t)x(t+T)>=\int dy dx y x f_2(y,t+T;x,t)$, require a two point density $f_2(y,t+T;x,t)$ for their description and provide us with limited information about the class of dynamics under consideration.

Two point conditional probability densities $p_k$, or transition probability densities, can then be defined as [5,6]:

$$f_2(x_1,t_1;x_1,t_1) = p_2(x_2,t_2|x_1,t_1)f_1(x_1,t_1), \ (1)$$

$$f_3(x_3,t_3;x_2,t_2;x_1,t_1) = p_3(x_3,t_3|x_2,t_2,x_1,t_1)p_2(x_2,t_2|x_1,t_1)f_1(x_1,t_1)$$
, (2)

and more generally as

$$f_n(x_n,t_n;...;x_1,t_1) = p_n(x_n,t_n|x_{n-1},t_{n-1};...x_1,t_1)f_{n-1}(x_{n-1},t_{n-1};...;x_1,t_1)$$
$$= p_n(x_n,t_n|x_{n-1},t_{n-1};...x_1,t_1)...p_2(x_2,t_2|x_1,t_1)f_1(x_1,t_1)$$
, (3)

where $p_n$ is the 2-point conditional probability density to find $x_n$ at time $t_n$, given the last observed point $(x_{n-1}, t_{n-1})$ and the previous history $(x_{n-2}, t_{n-2}; \ldots; x_1, t_1)$. When memory is present in the system then one cannot use the simplest 2-point transition density $p_2$ to describe the complete time evolution of the dynamical system that generates $x(t)$.

In a Markov process the picture is much simpler. A Markov process [5,6] remembers only the last observed point in the time series. There, we have

$$f_n(x_n, t_n; \ldots; x_1, t_1) = p_2(x_n, t_n | x_{n-1}, t_{n-1}) \ldots p_2(x_2, t_2 | x_1, t_1) f_1(|x_1, t_1),$$
(4)

because all transition rates $p_n$, n>2, are built up as products of $p_2$,

$$p_k(x_k, t_k | x_{k-1}, t_{k-1}; \ldots; x_1, t_1) = p_2(x_k, t_k | x_{k-1}, t_{k-1}),$$
(5)

for k=3,4, .... , and so $p_2$ cannot depend on an initial state $(x_1, t_1)$ or on any previous state other than the last observed point $(x_{k-1}, t_{k-1})$. Only in the absence of memory does the 2-point density $p_2$ describe the complete time evolution of the dynamical system. E.g., we can prove that for an arbitrary process with or without memory

$$p_{k-1}(x_k, t_k | x_{k-2}, t_{k-2}; \ldots; x_1, t_1) = \int dx_{k-1} p_k(x_k, t_k | x_{k-1}, t_{k-1}; \ldots; x_1, t_1) p_{k-1}(x_{k-1}, t_{k-1} | x_{k-2}, t_{k-2}; \ldots; x_1, t_1)$$
(6)

and therefore that

$$p_2(x_3, t_3 | x_1, t_1) = \int dx_2 p_3(x_3, t_3 | x_2, t_2; x_1, t_1) p_2(x_2, t_2 | x_1, t_1),$$

(7)

whereas the Chapman-Kolmogorov (CK) equation for a Markov process follows with $p_n=p_2$ for n=2,3,…, from (6) so that

$$p_2(x_3,t_3|x_1,t_1) = \int dx_2\, p_2(x_3,t_3|x_2,t_2) p_2(x_2,t_2|x_1,t_1). \quad (8)$$

The Markov property is expressed by $p_n=p_2$ for all n≥3, the complete lack of memory excepting the last observed point. *The CK Equation (8) is a necessary but not sufficient condition for a Markov process [7,8,9,10].*

A time translationally invariant Markov process defines a 1-parameter semi-group $U(t_2,t_1)$ of transformations [10], where $p_2(x_n,t_n:x_{n-1},t_{n-1})=p_2(x_n,t_n-t_{n-1}:x_{n-1},0))$, but time translational invariance is not a property of FX data [4] and will not be assumed here. In any and all cases, the identity element is defined by the equal times transition density

$$p_2(y,t|x,t) = \delta(y-x). \quad (9)$$

Arbitrary processes with memory do not obey the CK eqn. Instead, the class of path-dependent time evolutions is defined by the entire hierarchy eqns. (3,6), for n=2,3,4,… . That the transition density for fBm, e.g., obeys no CK eqn. is shown implicitly in Appendix B of [11], where the authors show that for general Gaussian processes one obtains the semi-group property iff. the Gaussian describes a Markov process. However, both CK and Fokker-Planck eqns. have been shown to hold for Ito processes with only *finitely* many states in memory [9].

Memory-dependent processes in statistical physics have been discussed by Hänggi and Thomas [11]. They point our that

$$p_2(x_3,t_3|x_2,t_2) = \frac{\int dx_1 p_3(x_3,t_3|x_2,t_2;x_1,t_1)p_2(x_2,t_2|x_1,t_1)f_1(x_1,t_1)dx_1}{\int p_2(x_2,t_2|x_1,t_1)f_1(x_1,t_1)dx_1}$$
(10)

is a functionals of the initial state $f_1(x_1,t_1)$ in which the system was prepared at the initial time $t_1$, unless the process is Markovian. In a nonMarkov system one may sometimes be able to mask this dependence on state preparation by choosing the initial condition to be $f_1(x_1,t_1)=\delta(x_1)$. If, instead, we would or could choose $f_1(x_1,t_1)=\delta(x_1-x'_o)$ at $t_1=0$, e.g., then we obtain $p_2(x_3,t_3;x_2,t_2)=p_3(x_3,t_3;x_2,t_2,x_o')$, introducing a dependence on $x_o'$ in both the drift and diffusion coefficients. So in this case, what appears superficially as $p_2$ is really a special case of $p_3$. The authors of [11] point out that the origin of memory in statistical physics is often a consequence of averaging over other, rapidly changing, variables. We will mention memory as a consequence of averaging over other variables in the section below on the efficient market hypothesis.

A class of Markov processes with scaling more general than Hurst exponent scaling [1,3,12] is defined as follows: let

$$f_1(x,t) = \sigma_1^{-1}(t)F(u)$$  (11)

with initial condition $f_1(x,0)=\delta(x)$, where $u=x/\sigma_1(t)$, with variance[1]

$$\sigma^2(t) = \left\langle x^2(t) \right\rangle = \sigma_1^{2}(t)\left\langle u^2 \right\rangle.$$  (12)

---

[1] Mathematicians often write $x(t)=t^H x(1)$. The variable $x(1)$ is the same as our variable $u$: the time $t$ is here dimensionless, otherwise the diffusion coefficient $D(x,t)$ must be multiplied by a constant factor with dimension sec$^{-1}$.

Then with the diffusion coefficient scaling as

$$D(x,t) = (d\sigma_1^2 / dt)\overline{D}(u) \qquad (13)$$

where $d\sigma_1 / dt > 0$ is required, $f_1(x,t)$ satisfies the Fokker-Planck pde

$$\frac{\partial f_1}{\partial t} = \frac{1}{2}\frac{\partial^2 (Df_1)}{\partial x^2} \qquad (14)$$

and yields the scale invariant part of the solution

$$F(u) = \frac{C}{\overline{D}(u)}e^{-\int u du / \overline{D}(u)}. \qquad (15)$$

An example is given by Hurst exponent scaling $\sigma_1(t) = t^H$, $0 < H < 1$. A piecewise constant drift $R(t)$ can be included in our result by replacing $x$ by $x - \int R(s)ds$ in $u$ [1,12]. Note that the scaling (11,12) is more general than our Markov example (13-15).

The Green function $g(x,t;x_o,t_o)$ of (14) for an arbitrary initial condition $(x_o,t_o) \neq (0,0)$ does not scale [12], but then the 2-point transition density $p_2(x_2,t_2; x_1,t_1)$ for fBm does not scale either[1]. In all cases scaling, when it occurs, can only be seen in the special choice of conditional density $f_1(x,t) = p_2(x,t;0,t_o)$ with $t_o = 0$ for a Markov process.

The same 1-point density $f_1(x,t)$ may describe nonMarkovian processes because a 1-point density taken alone, without the

---

information provided by the transition densities, defines no specific stochastic process and may be generated by many different completely unrelated processses, including systems with long time increment autocorrelations like fBm [1]. We will show below that the 2-point transition density delineates fBM from a martingale, but that pair correlations cannot be used to distinguish an arbitrary martingale from a drift-free Markov process.

## 3. Absence of trend and martingales

By a trend, we mean that $d<x(t)>/dt \neq 0$, conversely, by lack of trend we mean that $d<x(t)>/dt=0$. If a stochastic process can be detrended, then $d<x>/dt=0$ is possible via a transformation of variables but one must generally specify which average is used to define $<x>$. If the drift coefficient $R(x,t)$ depends on $x$, then detrending with respect to a specific average generally will not produce a detrended series if a different average is then used (e.g., one can choose different conditional averages, or an absolute average). We next discuss processes that can be detrended once and for all by a simple subtraction. I.e., we assume for the time being a trivial drift coefficient but allow for nontrivial diffusion coefficients. This case is of interest both theoretically and for FX data analysis.

A trivial drift coefficient $R(t)$ is a function of time alone. A nontrivial drift coefficient $R(x,t)$ depends on $x$, on $(x,t)$, or on $(x,t)$ plus memory $\{x\}$, and is defined for Ito/Langevin processes by [5,6]

$$R(x,t,\{x\}) \approx \frac{1}{T} \int_{-\infty}^{\infty} dy(y-x)p_n(y,t+T;x,t,\{x\}) \quad (16)$$

as T vanishes, where {x} denotes the history dependence in $p_n$, e.g. with $y=x_n$ and $x=x_{n-1}$ $(y,t+T,x,t;\{x\})$ denotes $(x_n,t_n;x_{n-1},t_{n-1},x_{n-2},t_{n-2},\ldots,x_1,t_1)$ with $y=x_n$ and $x=x_{n-1}$. If $R(x,t)=0$ then

$$\int_{-\infty}^{\infty} dy \, y \, p_n(y,t+T;x,t,\{x\}) = x, \quad (17)$$

so that the conditional average over x at a later time is given by the last observed point in the time series, $<x(t+T)>_{cond}=x(t)$. This is the notion of a fair game: there is no systematic change in x on the average as t increases, $d<x(t)>_{cond}/dt=0$. The process x(t) is generally nonstationary, and the condition (17) is called a *local martingale* [13]. ***The possibility of vanishing trend, d<x>/dt=0, implies a local martingale x(t), and vice-versa.***

This is essentially the content of the Martingale Representation Theorem [14], which states that an arbitrary martingale can be built up from a Wiener process B(t), the most fundamental martingale, via stochastic integration ala Ito,

$$x(t) = \int b(x(s),s;\{x\})dB(s). \quad (18)$$

There is no drift term in (18). If a stochastic differential equation (sde)

$$dx(t) = b(x(t),t;\{x\})dB(t), \quad (19)$$

follows from (18) then the diffusion coefficient is defined by [5,6]

$$D(x,t,\{x\}) \approx \frac{1}{T} \int dy (y-x)^2 p_n(y,t+T|x,t,\{x\}) \quad (20)$$

as T vanishes, so that $D=b^2$. In a Markov system the drift and diffusion coefficients depend on (x,t) alone, have no history dependence. Ito calculus based on martingales has been developed systematically by Durrett, including the derivation of Girsanov's Theorem for arbitrary diffusion coefficients $D(x,t)$ [13]. Many discussions of Girsanov's Theorem [14,15] implicitly rule out the general case (19) where $D(x,t)$ may depend on x as well as t. In this paper we do not appeal to Girsanov's theorem because the emphasis is on application to data analysis, to detecting martingales in empirical data. A new and simplified proof of Girsanov's theorem for variable diffusion coefficients has been presented elsewhere [16].

There are stochastic processes that are inherently biased, and fBm provides an example. There, although the absolute average vanishes $<x(t)>=\int <x>_c f_1(x,t)dx=0$, the conditional average yields $d<x>_c/dt\neq0$ where the time dependence arises from long time correlations rather from a drift term: in fBm one obtains [1]

$$\left\langle x(t)\right\rangle_{cond} = \int dy y p_2(y,s|x,t) = C(t,s)x \qquad (21)$$

instead of the martingale condition (19). Here, $d<x>_c/dt=xdC/dt\neq0$ because the factor $C(t,s)\neq1$ is proportional to the autocorrelation function $<x(s)x(t)>$ where the stationarity of increments guaranteeing long time memory was built in [1]. Such processes cannot be 'detrended' ($R(x,t)=0$ by construction in fBm [1,17]) because what appears locally to be a trend in a conditional average is simply the strongly correlated behavior of the entire time series.

Note next that subtracting an average drift $\int$<R>dt from a process x(t) defined by x-dependent drift term plus a Martingale,

$$x(t) = x(t-T) + \int\limits_{t-T}^{t} R(x(s),s;\{x\})ds + \int b(x(s),s;\{x\}dB(s), \qquad (22)$$

does not produce a martingale. Here, if we replace x(t) by x(t)-$\int$<R>dt where the average drift term defined conditionally from some initial condition $(x_1, t_1)$,

$$\langle R \rangle = \int dx R(x,t) p_2(x,t|x_1,t_1) \qquad (23)$$

depends on t alone we do not get drift free motion, and choosing absolute or other averages of R will not change this. In financial analysis, e.g., <R> may represent an average from the opening return $x_1$ at opening time $t_1$ up to some arbitrary intraday return x at time t. The subtraction yields

$$x(t) = x(t-T) + \int\limits_{t-T}^{t} R(x(s),s;\{x\})ds - \int\limits_{t-T}^{t} \langle R \rangle ds + \int b(x(s),s;\{x\}dB(s)$$
(24)

and is not a martingale unless R is independent of x: we obtain <x(t)>$_c$=x iff. x=$x_o$. Constructing martingales for the special case of an (x,t) dependent drift R(x,t) in financial applications is carried out in part 6 below. Until part 6, we assume a trivial drift R(t) that has been subtracted, so that by x(t) we really mean x(t)-$\int$R(t)dt.

With that assumption we can, for our present purposes, divide stochastic processes into those that satisfy the martingale condition

$$\langle x(t) \rangle_{cond} = x(t_o), \qquad (25)$$

where $<..>_{cond}$ denotes the conditional average (19), and those that do not. Those that do not satisfy (19) can be classified further into processes that consist of a nontrivial (i.e., (x,t)-dependent) drift plus a martingale (20), and those (including fBm) that are not defined by an underlying martingale.

Finally, an Ito sde with or without drift included can be used to derive a Fokker-Planck pde for a stochastic process with memory. The Fokker-Planck pde is usually derived from the CK eqn. for a Markov process as an approximation, but this is not necessary. The derivation of the Fokker-Planck pde from the Ito sde, without assuming a C-K eqn. apriori, is provided in [9,19] goes through even if the drift and diffusion coefficients R and D are memory dependent. In that case one has a pde for a 2-point conditional probability $p_n$ depending on a history of n-2 earlier states.

## 4. Stationary vs. nonstationary increments

Let us preface this section with a comment: in contrast with what is assumed in the econophysics and finance literature, we know of only two stochastic processes with both finite variance and stationary increments: the Wiener process and fractional Brownian motion. Furthermore, we know of no finance data with stationary increments. Furthermore, stationary increments is a very different condition than either time or space translational invariance. Nonstationary increments are ubiquitous in both theory and data analysis.

In this section we generalize an argument in [1] that assumed Markov processes with trivially removable drift R(t). In fact, that argument was based on no specifically Markovian assumption and applies quite generally to nonMarkovian martingales. In the analysis that follows, we assume a drift-free nonstationary process x(t) with the initial condition $x(t_o)=0$, so that the variance is given by $\sigma^2=<x^2(t)>=\int x^2 f_1(x,t)$. By the increments of the process we mean x(t;T) = x(t+T)-x(t) and x(t;-T)=x(t)-x(t-T).

We state in advance that we assume that [$-\infty<x<\infty$}, that there are no boundary conditions that would lead to statistical equilibrium. All processes considered are nonstationary ones.

Stationary increments are defined by

$$x(t + T) - x(t) = x(T), \qquad (27)$$

'in distribution', and by nonstationary increments [1,3,4,5] we mean that

$$x(t + T) - x(t) \neq x(T). \qquad (28)$$

in distribution. When (27) holds, then given the density of 'positions' $f_1(x,t)$, we also know the density $f_1(x(T),T)=f_1(x(t+T)-x(t),T)$ of increments independently of the starting time t. Whenever the increments are nonstationary then any analysis of the increments inherently requires the two-point density, $f_2(x(t+T),t+T;x(t),t)$. From the standpoint of theory there exists no 1-point density of increments f(x;T),T) depending on T alone, independent of t, and spurious 1-point histograms of increments are typically constructed empirically by assuming that the converse is possible [4]. Next, we place an important restriction on the class of stochastic processes under consideration.

According to Mandelbrot, so-called 'efficient market' has no memory that can be *easily* exploited in trading [18]. Beginning with that idea we can assert the necessary but not sufficient condition, the absence of increment autocorrelations,

$$\left\langle (x(t_1) - x(t_1 - T_1))(x(t_2 + T_2) - x(t_2)) \right\rangle = 0, \qquad (29)$$

when there is no time interval overlap, $t_1 < t_2$ and $T_1$, $T_2 > 0$. This is a much weaker condition and far more interesting than asserting that the increments are statistically independent. We will see that this condition leaves the question of the dynamics of x(t) open, except to rule out processes with increment autocorrelations, specifically stationary increment processes like fBm [1,20], but also processes with correlated nonstationary increments like the time translationally invariant Gaussian transition densities described in [2].

Consider a stochastic process x(t) where the increments (29) are uncorrelated. From this condition we easily obtain the autocorrelation function for positions (returns), sometimes called 'serial autocorrelations'. If t>s then

$$\left\langle x(t)x(s) \right\rangle = \left\langle (x(t) - x(s))x(s) \right\rangle + \left\langle x^2(s) \right\rangle = \left\langle x^2(s) \right\rangle > 0, \quad (30)$$

since with $x(t_o)=0$ $x(s)-x(t_o)=x(s)$, so that $<x(s)x(t)>=<x^2(s)>$ is simply the variance in x. Given a history (x(t), …,x(s),…,x(0)), or (x(t_n),…x(t_k),…,x(t_1)), (30) reflects a martingale property:

$$\left\langle x(t_n)x(t_k) \right\rangle = \int dx_n...dx_1 x_n x_k p_n(x_n, t_n | x_n, t_n, ..., x_n, t_n, ...)p_{n-1}(...)...p_{k+1}(...)f_k(...)$$

$$= \int x_k^2 f_k(x_k, t_k; ...; x_1, t_1)dx_k...dx_1 = \int x^2 f_1(x,t)dx = \left\langle x_k^2(t_k) \right\rangle$$
$$(31)$$

where

$$\int x_m dx_m p_m(x_m, t_m | x_{m-1}, t_{m-1}; ...; x_1, t_1) = x_{m-1}. \qquad (32)$$

***Every martingale generates uncorrelated increments and conversely, and so for a Martingale <x(t)x(s)>=<x²(s)> if s<t.²***

In a martingale process, the history dependence cannot be detected at the level of 2-point correlations, memory effects can at best first appear at the level 3-point correlations requiring the study of a transition density $p_3$. Here, we have not postulated a martingale, instead we've deduced that property from the lack of pair wise increment correlations. But this is only part of the story. What follows next is crucial for avoiding mistakes in data analysis [4].

Combining

$$\left\langle (x(t+T) - x(t))^2 \right\rangle = +\left\langle \left\langle x^2(t+T) \right\rangle + \left\langle x^2(t) \right\rangle - 2\left\langle x(t+T)x(t) \right\rangle \right.$$
(33)

with (34), we get

$$\left\langle (x(t+T) - x(t))^2 \right\rangle = \left\langle x^2(t+T) \right\rangle - \left\langle x^2(t) \right\rangle \qquad (34)$$

which depends on *both* t and T, excepting the case where <x²(t)> is linear in t. Uncorrelated increments are generally nonstationary. *Therefore, martingales generate uncorrelated, typically nonstationary increments.* So, at the level of pair correlations a martingale with memory cannot be distinguished empirically from a drift-free Markov process. *To see the memory in a martingale one must study at the very least the 3-point correlations.*

---

² Note that (30,31) hold for time translationally invariant martingales, where $p_2(x,t:y,s)=p_2(x,t-s:y,0)$. One can easily check this for a drift-free Gaussian Markov process. I.e., time translational invariance does not imply that <x(t)x(s)> is a function of t-s alone. Time translational invariance of $p_n$, n≥2, does not imply that a statistical equilibrium density $f_1(x)$ exists *and is approached asymptotically* by $f_1(x,t)$ [21]. I.e., a time translationally invariant martingale on [-∞,∞] cannot yield a stationary process, cannot lead to statistical equilibrium.

Summarizing, we've shown explicitly that fBm is not a martingale [1], while every Markov process with trivial drift R(t) can be transformed into a (local) Martingale via the substitution of x(t)-∫Rdt for x(t): Ito sdes with vanishing drift describe local martingales [13]. A martingale may have memory, and we've provided a model diffusion coefficient to illustrate the appearance of memory (any drift or diffusion coefficient depending on a state (x',t') other than the present state (x,t) exhibits memory, so diffusive models with memory are quite easy to construct). We've shown that uncorrelated increments are nonstationary unless the variance is linear in t. This means that looking for memory in two point correlations is useless: at that level of description a martingale with memory will look Markovian. To find the memory in a martingale one must study the transition densities $p_n$ and correlations for n≥3. This has not been discussed in the literature, so far as we know.

## 5. The Efficient Market Hypothesis

We begin by sumarizing our viewpoint for the reader. Real finance markets are hard to beat, arbitrage posibilites are hard to find and, once found, tend to disappear fast. In our opinion the EMH is simply an attempt to mathematize the idea that the market is very hard to beat. If there is no useful information in market prices, then those prices can be counted as noise, the product of 'noise trading'. A martingale formulation of the EMH embodies the idea that the market is hard to beat, is overwhelmingly noise, but leaves open the question of hard to find correlations that might be exploited for exceptional profit.

A strict interpretation of the EMH is that there are no correlations, no patterns of any kind, that can be employed *systematically* to beat the average return <R> reflecting the market itself: if one wants a higher return, then one must take on more risk. A Markov market is unbeatable, it has no systematically repeated patterns, no memory to exploit. We will argue below that the stipulation should be added that in discussing the EMH we should consider only normal, liquid markets, meaning very liquid markets with small enough transactions that approximately reversible trading is possible on a time scale of seconds [3]. Otherwise, 'Brownian' market models do not apply. Liquidity, 'the money bath' created by the noise traders whose behavior is reflected in the diffusion coefficient [3], is somewhat qualitatively analogous to the idea of the heat bath in thermodynamics [24]: the second by second fluctuations in x(t) are created by the continual noise trading.

Mandelbrot [18] proposed a less strict and very attractive definition of the EMH, one that directly reflects the fact that financial markets are hard to beat but leaves open the question whether the market can be beaten in principle at some high level of insight. He suggested that a martingale condition on returns realistically reflects the notion of the EMH. A martingale may contain memory, but that memory can't be easily exploited to beat the market precisely because the expectation of a martingale process x(t) at any later time is simply the last observed return. In addition, as we've shown above, pair correlations in increments cannot be exploited to beat the market either. The idea that memory may arise (in commodities, e,g.) from other variables (like the weather) [18] correponds in statistical physics [11] to the appearance of memory as a consequence of averaging over other, more rapidly changing, variables in the larger dynamical system.

The martingale (as opposed to Markov) version of the EMH is also interesting because technical traders assume that certain price sequences give signals either to sell or buy. In principle, that is permitted in a martingale. A particular price sequence $(p(t_n), \ldots, p(t_1))$, were it quasi-systematically to repeat, can be encoded as returns $(x_n, \ldots, x_1)$ so that a conditional probability density $p_n(x_n; x_{n-1}, \ldots, x_1)$ could be interpreted as a providing a risk measure to buy or sell. By 'quasi-repetition' of the sequence we mean that $p_n(x_n; x_{n-1}, \ldots, x_1)$ is significantly greater than a Markovian prediction. Typically, technical traders make the mistake of trying to interpret random price sequences quasi-deterministically, which differs from our interpretation of 'technical trading' based on conditional probabilities (see Lo et al [25] for a discussion of technical trading claims, but based on a non-martingale, non-empirically based model of prices). With only a conditional probability for 'signaling' a specific price sequence, an agent with a large debt to equity ratio can easily suffer the Gamblers' Ruin. In any case, we can offer no advice about technical trading, because the existence of market memory has not been firmly established (the question is left open by the analysis of ref. [25]), liquid finance markets look pretty Markovian so far as we've been able to understand the data [4], but one can go systematically beyond the level of pair correlations to try to find memory. Apparently, this remains to be done, or at least to be published.

Memory could reflect heavy trading around a particular price and can, of course, be lost in the course of time. The writer remembers well the period of a few months ca. 1999 when CPQ sold for around $22, and was traded often in the range $18-$25 before crashing further. Whether that provides an example is purely speculation at this point.

Fama [26] took Mandelbrot's proposal seriously and tried to test finance data at the simplest level for a fair game condition. We continue our discussion by first correcting a mathematical mistake made by Fama (see the first two of three unnumbered equations at the bottom of pg. 391 in [26]), who wrongly concluded in his discussion of martingales as a fair game condition that $<x(t+T)x(t)>=0$. Here's his argument, rewritten partly in our notation. Let $x(t)$ denote a 'fair game'. With the initial condition chosen as $x(t_o)=0$, then we have the unconditioned expectation $<x(t)>=\int x dx f_1(x,t)=0$ (there is no drift). Then the so-called 'serial covariance' is given by

$$\langle x(t+T)x(t) \rangle = \int x dx < x(t+T) >_{cond(x)} f_1(x,t). \qquad (40)$$

Fama states that this vanishes because $<x(t+T)>_{cond}=0$. This is impossible: by a fair game we mean a Martingale, the conditional expectation is $<x(t+T)>_{cond}=\int y dy p_2(y,t+T;x,t)=x=x(t)\neq 0$, and so Fama should have concluded instead that $<x(t+T)x(t)>=<x^2(t)>$ as we showed in the last section. Vanishing of (40) would be true of statistically independent variables but is violated by a 'fair game'. Can Fama's argument be salvaged? Suppose that instead of $x(t)$ we would try to use the *increment* $x(t,T)=x(t+T)-x(t)$ as variable. Then $<x(t,T)x(t)>=0$ for a Martingale, as we showed in part 4. However, Fama's argument still would not be generally correct because $x(t,T)$ canno*t* be taken as a 'fair game' variable unless the variance is linear in t, and in financial markets the variance is not linear in t [3,4]. Fama's mislabeling of time dependent averages (typical in economics and finance literature) as 'market equilibrium' has been corrected elsewhere [24].

In our discussion of the EMH we shall not follow the economists' tradition and discuss three separate forms (weak, semi-strong, and strong [27]) of the EMH, where a hard to test or effectively nonfalsifiable distinction is made between three separate classes of traders. We specifically consider only normal liquid markets with trading times at multiples of 10 min. intervals so that a Martingale condition holds [4]. Normal market statistics overwhelmingly (with high probability, if not 'with measure one') reflect the noise traders [3], so we consider *only* normal liquid markets and ask whether noise traders produce signals that one might be able to trade on systematically. The question whether insiders, or exceptional traders like Buffett and Soros, can beat the market probably cannot be tested scientifically: even if we had statistics on such exceptional traders, those statistics would likely be too sparse to draw a firm conclusion (see [3,4] for a discussion of the difficulty of getting good enough statistics on the noise traders, who dominate a normal market). Furthrmore, it is not clear that they beat liquid markets, some degree of illiquidity seems to play a significant role there. Effectively, or with high probability, there is only one type trader under consideration here, the noise trader. Noise traders provide the liquidity [28], their trading determines the form of the diffusion coefficient D(x,t;{x}) [3], where {x} reflects any memory present. The question that we emphasize is whether, given a Martingale created by the noise traders, a normal liquid market can still be beaten systematically.

Testing the market for a nonMarkovian martingale is nontrivial and apparently has not been done: tests at the level of pair correlations leave open the question of higher order correlations that may be exploited in trading. Whether the hypothesis of a martingale as EMH will stand the test of higher orders correlations exhibiting memory remains to be seen. In the long run, one may be required to identify a very liquid 'efficient market' as Markovian.

Finally, martingales typically generate nonstationary increments. This means that it is generally impossible to use the increment x(t,T) (or the price difference p(t+T)-p(t)) as a variable in the description of the underlying dynamics. The use of a returns or price increment as variable in data analysis generates spurious Hurst exponents [4,31] and spurious fat tails whenever the time series have nonstationary increments [3,4]. The reason that an increment cannot serve as a 'good' coordinate is that it depends on the staring time t: let z=x(t;T). Then

$$f(z,t,t+T) = \int f_2(y,t+T;x,t)\delta(z-y+x)dxdy \qquad (41)$$

is not independent of t, although attempts to construct this quantity as histograms in data analysis via 'sliding windows' implicitly presume t-independence [4,31]. If the increments are stationary then z=y-x=x(T) and we obtain a well defined density f(z,T). When the increments are nonstationary then f depends on t and (42) can be seen a failed attempt to coarsegrain.

## 7. Levy's definition of Brownian motion, a cautionary note

Levy's characterization of "Brownian motion" (meaning the Wiener process) is stated in various equivalent ways in the literature (pg. 46 in Friedman [10], pg. 75 Durrett [13], pg. 204 in Steele [14], and pg. 111 in Durrett [34]) We can identify the careless reading of that theorem as the source of the false expectation expressed in much of the finance literature that an arbitrary martingale is equivalent by a change of time variable to a Wiener process (see pg. 204-5 in Steele [14] for that mistake, but see also pg. 75 in [13] for the same claim). Levy's definition can be stated as follows [10]: with the assumptions that Y(t) and $Y^2(t)$-t are both martingales, then Y(t) is a Wiener process within a change of time variable. Here's the most general construction of a martingale from Ito calculus: let x(t) be any Ito process dx=R(x,t)dt+√D(x,t)dB(t). A local martingale Y(t)=G(x,t) can be constructed by setting the drift term equal to zero in Ito's lemma (requiring that G(X,t) satisfies Kolmogorov's backward time pde subject to initial and boundary conditions) and is generated by the sde

$$dY = \frac{\partial G}{\partial x} \sqrt{D(x,t)} dB. \qquad (47)$$

Durrett [13] shows how to construct and do Ito calculus with martingales, a generalization of the standard case where Ito differentials and stochastic integration are developed for Wiener processes [10,13,14,15]. For a martingale Y, the easy to derive integration by parts formula becomes [13]

$$Y(t) - Y(t_o) = \int (dY)^2 + 2\int Y dY, \qquad (48)$$

where $(dY)^2 = E(x,t)dt$ with $E(x,t) = G'^2(x,t)D(x,t)$, showing that $Y^2(t) - \int (dY)^2$ is a martingale. This reduces to the Wiener martingale $Y^2(t) - t$ iff. $G'(x,t)\sqrt{D(x,t)} = 1$. E.g., for the drift-free exponential process [9] with $H=1/2$ and $x(0)=0$, $<x^2(t)>=2t$, showing that $<x^2(t)>-2t$ is a martingale, and therefore $<x^2(t)>-t$ is not.

Durrett [13] emphasizes continuity of paths in his discussion of Levy's theorem. Scaling Markov processes [9] processes are generated by a drift-free sde with by $D(x,t) = |t|^{2H-1} D(u)$ where $u = |x| / |t|^H$, and satisfy the required conditions [21] for uniqueness and continuity of paths x(t) if the diffusion is not stronger than quadratic, $D(u) = 1+u^n$ with $n \leq 2$, and if $t>0$. The restriction to $t>0$ would seem problematic, but we've shown by direct construction [9] that Green functions g(x,t:y,s), $g(x,t:y,t) = \delta(x-y)$, exist for those processes at least for y=0, s=0, independent of n, so long as the indefinite integral ∫udu/D(u) is finite. The exponential process is generated by D(u)=1+u.

To complete the proof, we can show that the integrability requirements for the transformation of an arbitrary martingale X(t), dX=√D(X,t)dB(t), to a Wiener process B(t) are not satisfied. Assume a transformation Y(t)=G(X,t) such that dY=μ(t)dt+σ(t)dB, i.e., Y is to be a time change on a Wiener process with drift, where σ(t)≠1 defines a time change on the Wiener process. From Ito's lemma we obtain

$$\frac{\partial G}{\partial X}\sqrt{D(X,t)} = \sigma(t)$$
$$\frac{\partial G}{\partial t}+\frac{D(X,t)}{2}\frac{\partial^2 G}{\partial X^2} = \mu(t)$$
, (49)

and therefore

$$\frac{\partial G}{\partial t} = \mu(t)+\frac{\sigma(t)}{4}D^{-1/2}\frac{\partial D}{\partial X}. \qquad (50)$$

The integrability condition

$$\frac{\partial^2 G}{\partial t\partial X} = \frac{\partial^2 G}{\partial X\partial t} \qquad (51)$$

then yields

$$\frac{d\sigma}{dt} = \sigma[\frac{\partial D/\partial t}{2D}+\frac{1}{4}\frac{\partial^2 D}{\partial X^2}-\frac{(\partial D/\partial X)^2}{8D}] = 0. \qquad (52)$$

*With D(X,t) specified in advance, this equation produces a factor σ(t) independendent of X iff. D(X,t) is independent of X.* In that case

$$\sigma(t) = C\sqrt{D(t)} \qquad (53)$$

yields merely a time change on standard Brownian motion B(t) (meaning the Wiener process). Steele (pg. 205 in [15]) explicitly and apparently unknowingly restricts himself to this case, and the discusion of Girsanov's theorem in references [14,15] is also restricted to this case by virtue of the assumption that adding a drift term R to a Wiener process yields another Wiener process (that is possible iff. the drift coefficient R is independent of x!). As Durrett [13] shows while using notation that is misleading for a physicist ("<X>"is not an average of X but rather means ∫(dX)²=∫E(X,t)dt, e.g.), the correct statement of the Girsanov theorem is that removing an arbitrary drift term A via the Cameron-Martin-Girsanov transformation from a martingale X(t) plus the drift A, X(t)+A, yields another martingale M(t), and we see clearly that, in general, is neither of these martingales a Wiener process. "Intrinsic time" of the sort assumed by Durrett and Steele is discussed *explicitly* for the case where the diffusion coeficient D(t) depends on t alone by McKean (pg. 29 in [34]). The idea of 'intrinsic time', a special time variable where H=1/2 so that increments are stationary, is constructed locally by Gallucio et al in an empirical analysis [31], where we know that the diffusion coefficient depends on both x and t [4].

If we ask which time translationally invariant diffusions, dX=√D(X)dB, map to a Wiener process, then (53) yields

$$-D\frac{\partial^2 D}{\partial X^2} + \frac{(\partial D/\partial X)^2}{2} = cD \qquad (54)$$

with c a constant. This pde has at least one solution, $D(X)=aX^2$ with a>0 a constant and c=0. We obtain the transformation Y=lnX mapping the lognormal process X(t) to the Wiener process Y(t) =-(a/2)t+√aB(t). So Wiener processes with different time scales map to wiener processes, and the lognormal process maps to a wiener process. Aside from those special cases, the pot is empty.

Sumarizing, we've shown that *arbitrary martingales are topologically inequivalent to Wiener processes*: there is no global transformation Y=G(X,t) of an arbitrary martingale X to a Wiener process. This is analogous to nonintegrability in deterministic nonlinear dynamics, where chaotic and complex motions are topologically inequivalent to globally integrable ones. Locally, every Ito process reduces to a Wiener process with drift, and this is analogous to local integrability in dynamical systems theory where all deterministic motions satisfying a Lifshitz condition can be mapped locally to translations at constant speed on a lower dimensional manifold [34,35]. Assuming in the literature that arbitrary martingales are equivalent to Wiener processes trivializes martingales, and also leads to mistakes in calculations of first passage times, or 'hitting times': having provided us with the correct general formalism for stochastic calculus based on martingales, Durrett (eqn. (1.5) on pg. 212 of [36]) wrongly assumes with no explanation that Levy's theorem guarantees that an arbitrary martingale is merely a time transformation on a Wiener process.

## Acknowledgement

Kevin E. Bassler is supported by the NSF through grants #DMR-0406323 and #DMR-0427938. Gemunu H. Gunaratne is supported by the NSF through grant #DMS-0607345 and by TLCC. Joseph L. McCauley is grateful to a referee of [1] for encouraging us to extend our analysis to include the EMH, to Harry Thomas for sending us four of his papers on the time evolution of nonMarkovian systems and to extremely helpful criticism, comments, and suggested corrections via email, and to Enrico Scalas for pointing us to Doob's assertion that a Chapman-Kolmogorov Equation is not a sufficient condition to make a process Markovian [32], which led us to ref. [7,8]. JMC is also especially grateful to Giulio Bottazzi, who in an email discussion [37] suggested integrability conditions similar to those of part 7 by attempting to construct a transformation from an arbitrary Ito process with drift R(x,t) to a Wiener process with $\mu(t)=0$ and $\sigma(t)=1$. Our good friend and colleague Harry Thomas always careful and always helpful, pointed out the error on the initial condition of fBm in ref. [1], which we corrected in this paper, and also found two mistakes in an earlier version of this ms.